\documentclass[aps,prc,twocolumn,superscriptaddress,showpacs,floatfix,nofootinbib]{revtex4-1}
\usepackage{amsmath,graphicx}
\newcommand{\avgev}[1]{\left\langle{#1}\right\rangle}
\begin{document}

\title{Characterizing flow fluctuations with moments}

\author{Rajeev S. Bhalerao}
\affiliation{Department of Theoretical Physics,
Tata Institute of Fundamental Research,
   Homi Bhabha Road, Colaba, Mumbai 400 005, India}
\author{Jean-Yves Ollitrault}
\affiliation{
CNRS, URA2306, IPhT, Institut de physique th\'eorique de Saclay, F-91191
Gif-sur-Yvette, France} 
\author{Subrata Pal}
\affiliation{Department of Nuclear and Atomic Physics,
Tata Institute of Fundamental Research, Homi Bhabha Road, Mumbai, 400005, India}
\date{\today}

\begin{abstract}
We present a complete set of multiparticle correlation observables for
ultrarelativistic heavy-ion collisions. These include moments of the
distribution of the anisotropic flow in a single harmonic, and also
mixed moments, which contain the information on correlations between
event planes of different harmonics. We explain how all these moments
can be measured using just two symmetric subevents separated by a
rapidity gap. This presents a multi-pronged probe of the physics of
flow fluctuations. For instance, it allows to test the hypothesis
that event-plane correlations are generated by non-linear hydrodynamic
response. We illustrate the method with simulations of events in A
MultiPhase Transport (AMPT) model.
\end{abstract}

\pacs{25.75.Ld, 24.10.Nz}

\maketitle

\section{Introduction}
Large anisotropic flow has been observed in ultra-relativistic 
nucleus-nucleus collisions at the Relativistic Heavy-Ion Collider (RHIC) and
the Large Hadron Collider (LHC) ~\cite{Heinz:2013th}.  
Anisotropic flow is
an azimuthal ($\varphi$) asymmetry of the single-particle distribution~\cite{Luzum:2011mm}:
\begin{equation}
\label{defVn}
P(\varphi)=\frac{1}{2\pi}\sum_{n=-\infty}^{+\infty}V_n e^{-in\varphi},
\end{equation}
where $V_n$ is the (complex) anisotropic flow coefficient in the $n$th
harmonic.
One usually uses the notation $v_n$ for the magnitude: $v_n\equiv|V_n|$. 
Anisotropic flow is understood as 
the hydrodynamic response to spatial deformation of the initial density profile. 
This profile fluctuates event to event, which implies that the flow also 
fluctuates~\cite{Miller:2003kd,Alver:2006wh}. The recognition of the importance of 
flow fluctuations has led to a wealth of new flow observables, among which triangular 
flow~\cite{Alver:2010gr} and higher harmonics, as well as correlations between 
different Fourier harmonics~\cite{Aad:2014fla}. 

Flow fluctuations provide a window~\cite{Luzum:2013yya} into both the
early stage dynamics 
and the transport properties of the quark-gluon plasma. 
Specifically, the magnitudes of higher-order harmonics ($V_3$ to
$V_6$) are increasingly sensitive to the shear viscosity to entropy density
ratio~\cite{Luzum:2012wu}.
The distributions of $V_2$ and $V_3$ carry detailed information about
the initial density profile~\cite{Renk:2014jja,Yan:2014nsa}, while 
$V_4$ and higher harmonics are understood as superpositions of linear
and nonlinear responses, through which they are correlated with
lower-order harmonics~\cite{Teaney:2012ke,Bravina:2013ora}. 
Ideally, one would like to measure the full probability 
distribution $p(V_1,V_2,\cdots,V_n)$~\cite{Jia:2014jca}. 
So far, only limited information has been obtained, concerning 
either the distribution of a single $V_n$~\cite{Aad:2013xma} or
specific angular correlations between different
harmonics~\cite{Aad:2014fla}. 

We propose to study the distribution $p(V_1,V_2,\cdots,V_n)$ 
via its moments in various
harmonics~\cite{Bilandzic:2013kga,Bilandzic:2014qga}, either single or
mixed, and illustrate our point with realistic simulations 
using the AMPT model~\cite{Lin:2004en}. 
In Sec.~\ref{s:Method}, we recall how moments can be measured 
simply with a single rapidity gap~\cite{Bhalerao:2013ina}. 
This procedure is less demanding in terms of detector acceptance than the one
based on several rapidity windows separated pairwise by gaps 
\cite{Aad:2014fla}, and can be used to study even four-plane correlators.
In Sec.~\ref{s:old}, we list standard measures of flow
fluctuations which have been used in the literature and express them in terms of moments. 
In Sec.~\ref{s:new}, we introduce new observables which 
shed additional light on the origin of event-plane correlations. 
For instance, a correlation between $(V_2)^2$ and $V_4$ has been observed, 
which increases with impact parameter~\cite{Aad:2014fla}. 
This correlation is usually understood~\cite{Teaney:2013dta} as an effect of the 
non-linear hydrodynamic response which creates a $V_4$ proportional to 
$(V_2)^2$~\cite{Borghini:2005kd,Gardim:2011xv,Teaney:2012ke}: 
the increase in the correlation is thus assumed to result from the increase of 
elliptic flow~\cite{Aamodt:2010pa}. 
We show that this hypothesis can be tested directly by studying 
how the correlation between $(V_2)^2$ and $V_4$ is correlated with the magnitude 
of $V_2$. 
We also investigate in a similar way the origin of the three-plane correlation 
between $V_2$, $V_3$ and $V_5$~\cite{Aad:2014fla}. 

\section{Measuring moments}
\label{s:Method}

The statistical properties of $V_n$ are contained in its moments, which 
are average values of products of $V_n$, of the form
\begin{equation}
\label{defmoments}
{\cal M}\equiv
\left\langle \prod_n {(V_n)^{k_n} (V_n^*)^{l_n}}\right\rangle,
\end{equation}
where $k_n$ and $l_n$ are integers, and angular brackets denote an average value over
events.
Note that $V_n^*=V_{-n}$ and $V_0=1$. 
Azimuthal symmetry implies that the only nontrivial moments satisfy~\cite{Bhalerao:2011yg}
\begin{equation}
\sum_n n k_n=\sum_n n l_n.
\end{equation}

We now describe a simple procedure for measuring these moments, which 
applies to harmonics $n\ge 2$, i.e., $k_1=l_1=0$. 
(We do not study here moments involving directed flow $V_1$~\cite{Bhalerao:2011yg}.)
We define in each collision the flow vector~\cite{Poskanzer:1998yz} 
by
\begin{equation}
\label{defqcomplex}
Q_n 
\equiv\frac{1}{N} \sum_j e^{in\varphi_j},
\end{equation} 
where the sum runs over $N$
particles seen in a reference detector, and $\varphi_j$ are their
azimuthal angles.\footnote{ 
The factor $1/N$ in Eq.~(\ref{defqcomplex}) means that we choose to
average over particles in each event~\cite{Luzum:2012da}, rather than 
summing~\cite{Poskanzer:1998yz,Bhalerao:2014mua} or dividing by
$1/\sqrt{N}$~\cite{Adler:2002pu}. This choice is 
discussed at the end of Sec.~\ref{s:old}.} 
One typically measures $Q_n$ in two different parts of the detector 
(``subevents''~\cite{Danielewicz:1985hn}) $A$ and $B$, 
which are symmetric around midrapidity and separated by a gap in 
pseudorapidity (i.e., polar angle)~\cite{Adler:2003kt}.  
The moment (\ref{defmoments}) is then given by 
\begin{equation}
\label{qmoments}
{\cal M}\equiv\left\langle \prod_n {(V_n)^{k_n} (V_n^*)^{l_n}}\right\rangle=
\left\langle \prod_n {(Q_{nA})^{k_n} (Q_{nB}^*)^{l_n}}\right\rangle,
\end{equation}
which one can symmetrize over $A$ and $B$ to decrease the statistical error. 
This configuration, with all factors of $Q_n$ on one side and all factors of 
$Q_n^*$ on the other side~\cite{Bhalerao:2013ina}, suppresses 
nonflow correlations and self correlations as long as only harmonics 
$n\ge 2$ are involved.
An alternative procedure, where self correlations are explicitly subtracted, is 
described in~\cite{Bilandzic:2013kga}. 

In order to illustrate the validity of the method, we perform
calculations using the AMPT model~\cite{Lin:2004en}. 
AMPT reproduces quite well
LHC data for anisotropic flow ($v_2$ to $v_6$) at all 
centralities~\cite{Han:2011iy,Xu:2011jm,Pal:2012tc}. 
The implementation adopted in this paper~\cite{Pal:2012gf} 
uses initial conditions from the HIJING 2.0
model~\cite{Deng:2010mv}, which contains nontrivial 
event-by-event fluctuations. 
Flow in AMPT is produced by elastic scatterings in the partonic
phase. In addition, the model contains resonance decays, and thus
nontrivial nonflow effects. In the present work,
subevent $A$ consists of all particles in the pseudorapidity range
$0.4<\eta<4.8$, and subevent $B$ is symmetric around mid-rapidity, 
so that there is an $\eta$ gap of $0.8$ between $A$ and $B$~\cite{Luzum:2010sp}.

The thumb rule for measuring moments is that smaller values of  $n$ are 
easier to measure because $v_n$ decreases with $n$ 
for $n\ge 2$. 
Lower order moments, corresponding to smaller values of $k_n$ and $l_n$,
are also easier because higher-order moments are plagued with large variances, 
which entail large statistical errors.

\section{$v_n$ fluctuations, event-plane correlations, standard candles}
\label{s:old}

\begin{figure}
 \includegraphics[width=.9\linewidth]{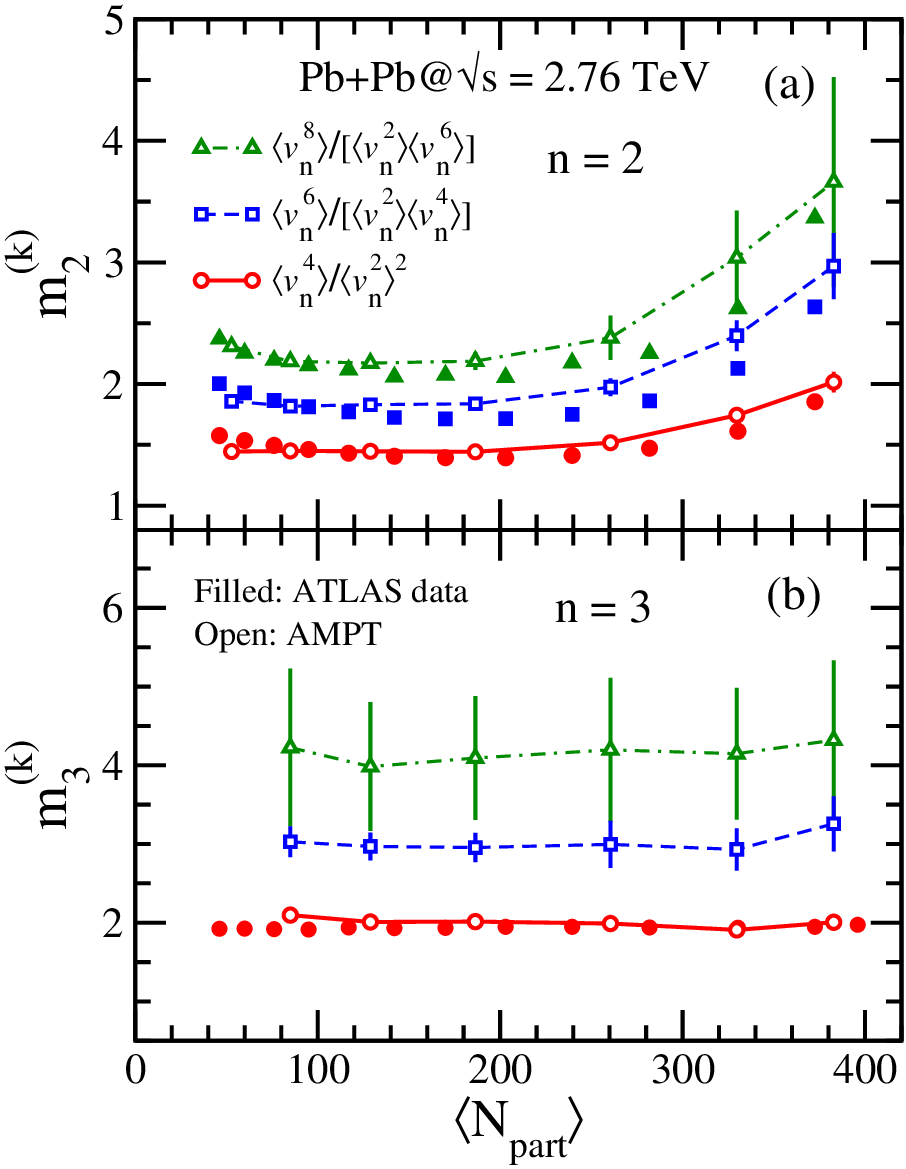}
 \caption{(Color online) Scaled moments of the distribution of $v_n$,
   (see Eq.~(\ref{scaledmoments})) for $k=2,3,4$, as a function of
   centrality, measured with the number of participant nucleons.
   Results are for (a) elliptic flow, $n=2$, and (b) triangular flow,
   $n=3$, in Pb-Pb collisions at $\sqrt{s}=2.76$ TeV. Open symbols
   represent AMPT calculations and closed symbols are obtained from
   ATLAS data \cite{Aad:2014vba}.}
\label{fig:scal}
\end{figure}

We first list observables which have been previously studied in the literature 
and explain how they can be measured using the method outlined in Sec.~\ref{s:Method}. 
Fluctuations of $v_n$ have been studied using 
cumulants~\cite{Borghini:2001vi,Adams:2004bi,Bilandzic:2010jr,Bzdak:2013rya}, 
which are linear combinations of even moments of the distribution of $v_n$, that is, 
$\langle (v_n)^{2k}\rangle$. 
These moments are obtained by keeping only one value of $n$ and 
setting $k_n=l_n=k$ in Eq.~(\ref{qmoments}).
Figure~\ref{fig:scal} displays the scaled moments
\begin{equation}
\label{scaledmoments}
m_n^{(k)}\equiv\frac{\langle v_n^{2k}\rangle}{\langle v_n^{2(k-1)}\rangle\langle v_n^{2}\rangle},
\end{equation}
for $k=2,3,4$ as a function of centrality for $n=2$ and $n=3$,
obtained by using the subevent method of Sec.~\ref{s:Method}.  
The scaled moment $m_n^{(k)}$ thus defined is invariant if one multiplies 
$v_n$ by a constant, therefore it reflects the statistics 
of $v_n$ and should be essentially independent of the detector acceptance. 
AMPT calculations are in fair agreement with the ATLAS data 
\cite{Aad:2014vba}, but tend to 
slightly overpredict $m_n^{(k)}$, i.e., overestimate flow 
fluctuations. 

If flow is solely created by fluctuations and if the statistics of these fluctuations is a 2-dimensional 
Gaussian~\cite{Voloshin:2007pc}, then $m_n^{(k)}=k$. 
As can be seen in Fig.~\ref{fig:scal}(b),  $m_3^{(k)}\simeq k$ for all centralities, 
as expected since $v_3$ is only from fluctuations in Pb-Pb 
collisions.\footnote{Small deviations from Gaussian statistics are actually seen experimentally 
and result in a non-zero cumulant $v_3\{4\}$~\cite{ALICE:2011ab}. Our simulation does not have enough 
statistics to detect this small non-Gaussianity.}
Similarly, as seen in Fig.~\ref{fig:scal}(a), $m_2^{(k)}$ is roughly equal to $k$ for central collisions where $v_2$ is mostly from Gaussian fluctuations, 
but decreases for mid-central collisions, 
corresponding to the emergence of a mean elliptic flow in the reaction plane~\cite{Ollitrault:1992bk}.

Event-plane correlations~\cite{Aad:2014fla} can also be expressed in
terms of moments which 
can be measured using the method outlined in Sec.~\ref{s:Method}, as already discussed in 
Ref.~\cite{Bhalerao:2013ina}. 
Specifically, two-plane correlations are Pearson correlation coefficients between 
moments.\footnote{Three- and four-plane~\cite{Jia:2012ju} correlations are not Pearson coefficients and 
are not bounded by unity~\cite{Luzum:2012da}.}
The Pearson correlation coefficient between two complex variables $f$ and $g$
whose average value is 0, $\avgev{f}= \avgev{g}=0$,
is defined as 
\begin{equation}
\label{Pearson}
r\equiv\frac{\avgev{fg^*}}{\sqrt{\avgev{|f|^2}\avgev{|g|^2}}}.
\end{equation}
$|r|\le 1$ in general, and $r=0$ if $f$ and $g$ are uncorrelated. 
The correlation between the second and fourth harmonic planes, which is denoted by 
$\avgev{\cos(4(\Phi_2-\Phi_4))}_w$ in Ref.~\cite{Aad:2014fla}, corresponds to 
$f=V_4$, $g=(V_2)^2$. 
The largest source of uncertainty in this measurement is the denominator which involves 
$\avgev{v_4^2}$, a measurement quadratic in the small harmonic $V_4$.

\begin{figure}
\includegraphics[width=.9\linewidth]{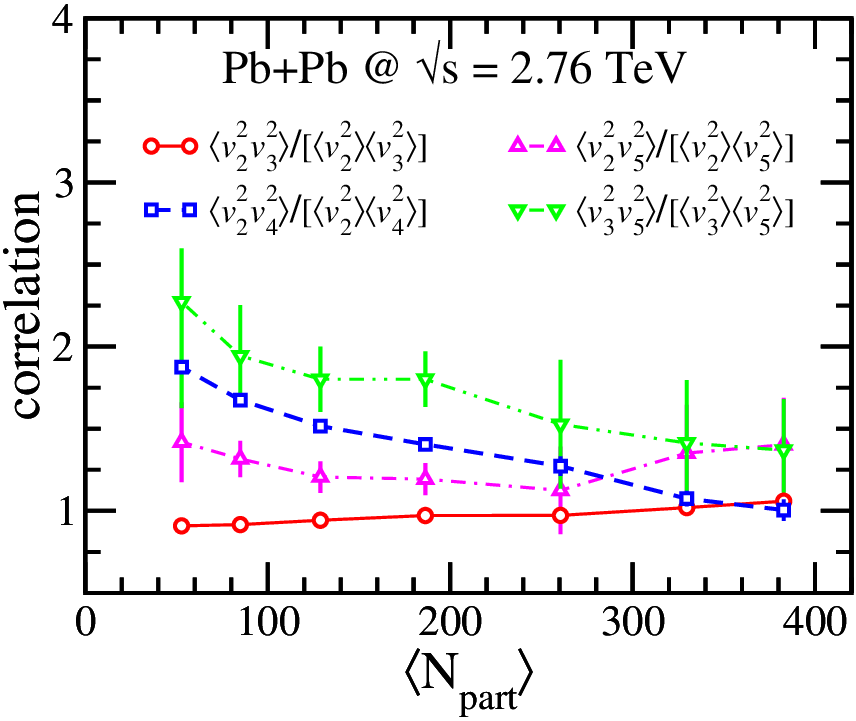}
\caption{(Color online) Correlations between $v_n^2$ and $v_m^2$
as a function of centrality for 
$(n,m)=(2,3)$, $(2,4)$, $(2,5)$, $(3,5)$, in Pb-Pb collisions at 
$\sqrt{s}=2.76$ TeV calculated in the AMPT model.}
\label{fig:standardcandles}
\end{figure}
Note that the scaled moments Eq.~(\ref{scaledmoments}) are
of the type $\avgev{fg}/\avgev{f}\avgev{g}$, which is another measure of the correlation 
between $f$ and $g$ when $\langle f\rangle$ and $\langle g\rangle$ both differ from 0. 
This correlation measure equals unity if $f$ and $g$ are uncorrelated, 
and is larger than unity if there is a positive correlation.
It is in general easier to measure than the Pearson correlation coefficient, 
because it does not involve the higher-order moments $\avgev{|f|^2}$ and $\avgev{|g|^2}$. 

The ``standard candles'' introduced in Ref.~\cite{Bilandzic:2013kga} correspond to the case
$f=v_n^2$, $g=v_m^2$, obtained by keeping only two harmonics, $n$ and $m$, and
setting $k_n=k_m=l_n=l_m=1$ in Eq.~(\ref{qmoments}).
These are correlations between the magnitudes $v_n$ and $v_m$, which do not involve 
the angular correlation between event planes.
Four of these correlations are displayed  in Fig.~\ref{fig:standardcandles}. 
The correlation between $v_2^2$ and $v_3^2$ is small and negative 
($\langle v_2^2v_3^2\rangle-\langle v_2^2\rangle\langle
v_3^2\rangle<0$), 
as already seen in AMPT calculations~\cite{Huo:2013qma}, 
while 
the correlation between the corresponding event planes is small and 
positive~\cite{Aad:2014fla}. 
All other correlations are positive. The correlation between $v_4$ and $v_2$,
and that between $v_5$ and $v_3$, become smaller for more central collisions, 
which is likely due to the smaller non-linear contributions~\cite{Gardim:2011xv} of $v_4$ and $v_5$, respectively.

Note that the observables in 
Eq.~(\ref{scaledmoments}) and Eq.~(\ref{Pearson}) are defined in
such a way that factors of $1/N$ in Eq.~(\ref{defqcomplex}) cancel
between the numerator and the denominator if $N$ is the same for all
events. 
In general, $N$ fluctuates, and the result depends on whether or not 
one includes a factor $1/N$. 
However, the centrality selection in experiments is typically done
using the multiplicity in a reference detector 
(see e.g. Ref.~\cite{Aamodt:2010pa}) 
so that effects of multiplicity fluctuations are likely to be small 
in narrow centrality intervals. 
The calculations in this paper are done with a $1/N$
normalization, but there is no strong argument for preferring one
normalization over another.

\section{Testing the non-linear response using moments}
\label{s:new}

\begin{figure}
 \includegraphics[width=\linewidth]{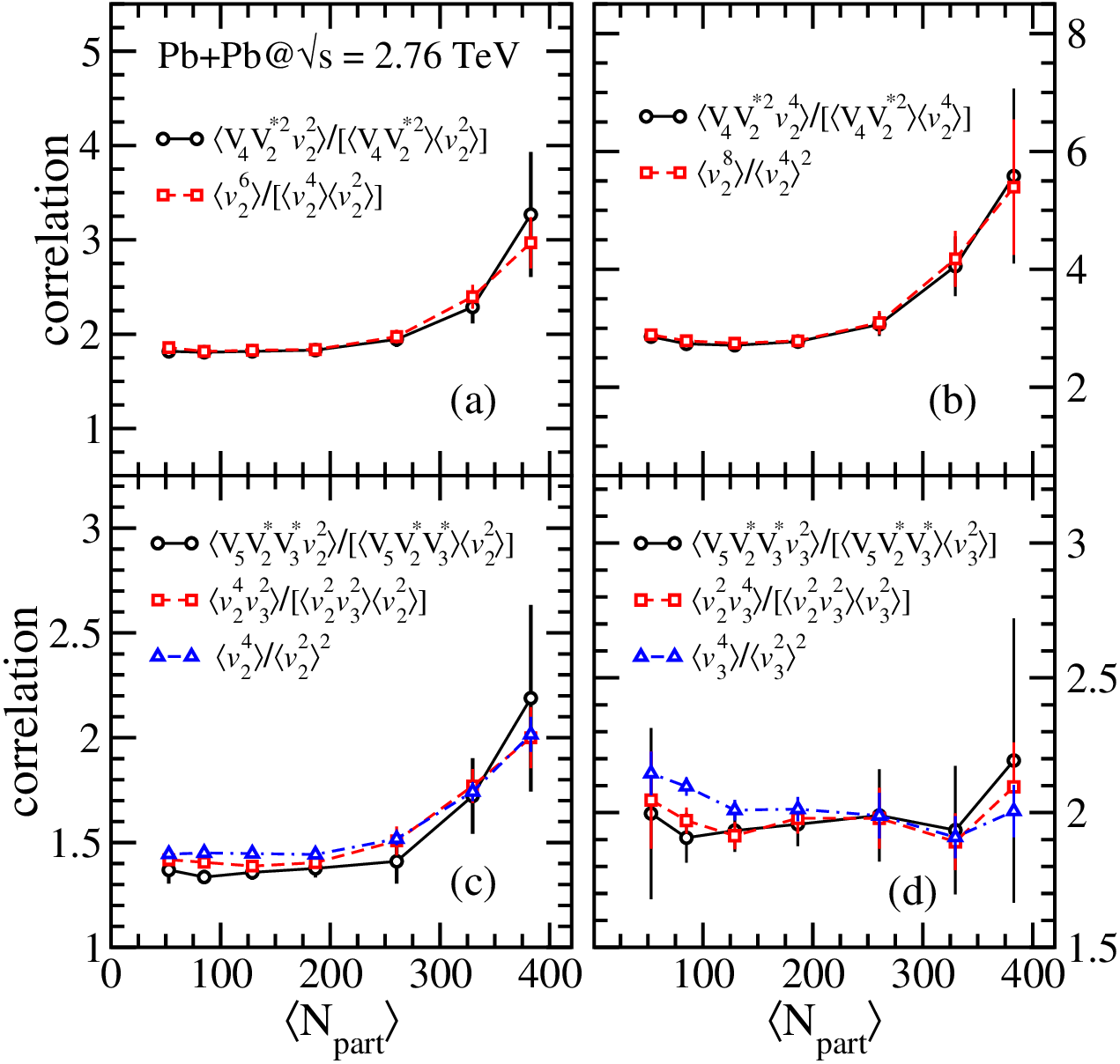}
 \caption{(Color online) Correlations between event-plane correlations and anisotropic flow calculated in AMPT,
and tests of Eqs.~(\ref{corrfluct1}), (\ref{corrfluct2}) and (\ref{corrfluct3}). Circles connected by solid lines correspond to the left-hand sides, and squares connected by dashed lines correspond to the right-hand sides of these equations.  }
\label{fig:all}
\end{figure}
We now introduce new correlation measures of the type $\avgev{fg}/\avgev{f}\avgev{g}$ 
in order to study how event-plane correlations are correlated with the magnitude
of anisotropic flow. 
We first consider the case $f=V_4(V_2^*)^2$ and $g=v_2^2$. 
$\avgev{fg}$ is obtained by setting $k_2=k_4=1$ and $l_2=3$ in Eq.~(\ref{qmoments}). 
The correlation $\avgev{fg}/\avgev{f}\avgev{g}$ is displayed in Fig.~\ref{fig:all}(a). 
There is a significant positive correlation for all centralities, which becomes larger 
for central collisions. 
In hydrodynamics, the correlation between $V_4$ and $(V_2)^2$ originates from a 
non-linear response~\cite{Teaney:2013dta}.  
In order to test this hypothesis, we model $V_4$ as the sum of two terms: 
 \begin{equation}
 \label{model}
 V_4=V_{4l}+\beta V_2^2, 
 \end{equation}
where the non-linear response coefficient 
$\beta$ is the same for all events in a centrality class.
This corresponds to the separation of $V_4$ into a linear part, created by fluctuations, 
and a non-linear part from $V_2$~\cite{Teaney:2012ke}. 
We assume in addition that $V_{4l}$ and $V_2$ are uncorrelated. 
Equation~(\ref{model}) then implies the following relation between moments:
\begin{equation}
\label{corrfluct1}
\frac{\avgev{V_4(V_2^*)^2 v_2^2}}{\avgev{V_4(V_2^*)^2}\avgev{v_2^2}}=
\frac{\avgev{v_2^6}}{\avgev{v_2^4}\avgev{v_2^2}}.
\end{equation}
This equation relates a mixed correlation between $V_4$  and $V_2$ to the fluctuations of 
elliptic flow: the right-hand side is the  scaled moment $m_2^{(3)}$ introduced in Eq.~(\ref{scaledmoments}). The
AMPT simulations support Eq.~(\ref{corrfluct1}) for all centralities, as can be seen in Fig.~\ref{fig:all}(a). 
A straightforward generalization of Eq.~(\ref{corrfluct1}) is obtained using $g=v_2^4$ instead 
of $v_2^2$:
\begin{equation}
\label{corrfluct2}
\frac{\avgev{V_4(V_2^*)^2 v_2^4}}{\avgev{V_4(V_2^*)^2}\avgev{v_2^4}}=
\frac{\avgev{v_2^8}}{\avgev{v_2^4}^2}.
\end{equation}
This correlation gives a higher weight to events with large elliptic flow. 
Equation~(\ref{corrfluct2}) is also supported by AMPT simulations, as shown in Fig.~\ref{fig:all}(b).

This discussion can be readily extended to the correlation between the 5th harmonic 
plane and the 2nd and 3rd harmonic planes.  
We now set $f=V_5V_2^*V_3^*$ and $g=v_2^2$ or $g=v_3^2$ and write 
\begin{equation}
\label{model5}
 V_5=V_{5l}+\beta' V_2V_3, 
 \end{equation}
where $V_{5l}$ is independent of $V_2$ and $V_3$ and $\beta'$ is constant. 
One thus obtains
\begin{eqnarray}
\label{corrfluct3}
\frac{\avgev{V_5V_2^*V_3^* v_2^2}}{\avgev{V_5V_2^*V_3^*}\avgev{v_2^2}}&=&
\frac{\avgev{v_2^4v_3^2}}{\avgev{v_2^2v_3^2}\avgev{v_2^2}},\cr
\frac{\avgev{V_5V_2^*V_3^* v_3^2}}{\avgev{V_5V_2^*V_3^*}\avgev{v_3^2}}&=&
\frac{\avgev{v_2^2v_3^4}}{\avgev{v_2^2v_3^2}\avgev{v_3^2}}.
\end{eqnarray}
The numerators in the right-hand side are obtained using Eq.~(\ref{qmoments}) with 
$k_2=l_2=2$, $k_3=l_3=1$ (first line) and  $k_2=l_2=1$, $k_3=l_3=2$ (second line). 
Figs.~\ref{fig:all}(c)-(d) again show that these equalities are very well verified by AMPT simulations. 
One can also use the fact that $v_2$ and $v_3$ are weakly correlated, as seen in 
Fig.~\ref{fig:standardcandles}, which leads to the following simplified relations: 
\begin{eqnarray}
\label{corrfluct4}
\frac{\avgev{v_2^4v_3^2}}{\avgev{v_2^2v_3^2}\avgev{v_2^2}}&\simeq& 
\frac{\avgev{v_2^4}}{\avgev{v_2^2}^2},\cr
\frac{\avgev{v_2^2v_3^4}}{\avgev{v_2^2v_3^2}\avgev{v_3^2}}&\simeq& 
\frac{\avgev{v_3^4}}{\avgev{v_3^2}^2}. 
\end{eqnarray}
These relations are also satisfied to a good approximation
(see Figs.~\ref{fig:all}(c)-(d)), thus showing that the 
correlators in Eq.~(\ref{corrfluct3}) are mostly driven by the fluctuations of 
$v_2$ and $v_3$.

\section{Conclusion}

Moments of the distribution of $V_n$ provide a complete set of
multiparticle correlation observables, which can be used to probe the physics of 
flow fluctuations in unprecedented detail.  
All these moments can be measured using just two subevents separated by a rapidity gap. 
Moments yield the full information on the multiparticle correlations
without resorting to unfolding procedures~\cite{Alver:2007qw,Aad:2013xma}
or event-shape engineering~\cite{Schukraft:2012ah,Dobrin:2012zx,Jia:2014jca}. 
They can be measured easily at LHC and even with detectors having
smaller acceptance, and can be
directly compared with theoretical calculations. 
In particular, scaled moments~\cite{Bhalerao:2011yg} as studied in this paper 
are typically independent of the details of the acceptance, and reflect global 
fluctuations. They can be used to further probe the physics of initial-state fluctuations 
and the hydrodynamic response of the quark-gluon plasma. 

We have shown that the assumption that the correlation between 
$V_4$ and $V_2$, and that between $V_5$, $V_2$ and $V_3$, are driven 
by nonlinear response, yields nontrivial relations between moments,
Eqs.~(\ref{corrfluct1}), (\ref{corrfluct2}) and (\ref{corrfluct3}). 
Simulations within the AMPT model have shown that these relations are
very well satisfied. 
It is important to test if experimental data confirm these
predictions. 
 
\begin{acknowledgments}
This work is funded by CEFIPRA under project 4404-2.
JYO thanks the MIT LNS for hospitality and 
acknowledges support by the European Research Council under the
Advanced Investigator Grant ERC-AD-267258. 
\end{acknowledgments}

\end{document}